\documentclass[letterpaper,twocolumn,10pt]{article}
\usepackage{usenix}

\usepackage{tikz}
\usepackage{amsmath}

\usepackage{filecontents}

\usepackage{amsmath}
\usepackage{amsthm}
\usepackage{etoolbox}
\usepackage{amssymb}
\usepackage{mathtools}
\usepackage{bm}
\usepackage{bbm}
\usepackage{booktabs}
\usepackage{xspace}
\usepackage{color,soul}
\usepackage{multirow}
\usepackage{multicol}
\usepackage{caption}
\usepackage{enumitem}
\usepackage{verbatim}
\usepackage{microtype}
\usepackage{graphicx}
\usepackage{booktabs}
\usepackage{hyperref}
\usepackage{nicefrac}
\usepackage[normalem]{ulem}
\usepackage{tikz}
\usetikzlibrary{trees}
\usepackage{float}
\usepackage{subcaption}
\usepackage[ruled,noend,linesnumbered,vlined]{algorithm2e}
\usepackage{xcolor,colortbl}
\usepackage{hhline}
\usepackage{tabularx}
\usepackage{listings}
\usepackage{minted}
\usepackage{xcolor}
\usepackage{makecell}
\usepackage{pifont}
\usepackage{stmaryrd}
\usepackage{fontawesome5}

\usepackage{array}
\newcolumntype{P}[1]{>{\raggedright\arraybackslash}p{#1}}

\usepackage{pifont}
\usepackage{lipsum}

\usepackage{tikz}

\newcommand{\intset}[1]{\llbracket#1\rrbracket}

\newcommand{\mysubsubsection}[1]{{\vspace{0.25em}\noindent{\textbf{#1.\xspace}}}}

\newcommand{\system}{{\text{Janus}}\xspace}
\newcommand{\systembase}{{\text{Janus\textsubscript{base}}}\xspace}

\makeatletter
\newcommand{\nakedthanks}[1]{%
  \protected@xdef\@thanks{%
    \@thanks
    \protect\@naked@thanks{#1}%
  }%
}
\newcommand{\@naked@thanks}[1]{%
  \let\thefootnote\relax
  \footnotetext{#1}%
}
\makeatother

\begin{document}

%don't want date printed
\date{}

\title{\Large \bf Efficient Agentic LLM Serving over SSD-based Sparse KV Storage}

\author{
\rm{Wenhao He}$^{\ast\ddag}$ \hspace{1.2em} 
\rm{Ping Zhang}$^{\ddag}$ \hspace{1.2em} 
\rm{Xiaohe Hu}$^{\S\ddag}$ \hspace{1.2em} 
\rm{Chutian Wang}  \\ 
\rm{Jinlong Hou}$^{\S}$ \hspace{1.2em} 
\rm{Yuan Cheng}$^{\S}$  \hspace{1.2em} 
\rm{Peng Sun}$^{\dag}$ \hspace{1.2em} 
\rm{Fangcheng Fu}$^{\ast\,\text{\faEnvelopeSquare}}$ 
 \\ [.6em] 
 $^{\ast}$Shanghai Jiao Tong University~~~$^{\ddag}$Infrawaves\\
$^{\S}$Shanghai Innovation Institute~~~$^{\dag}$Shanghai Qiji Zhifeng Co., Ltd.\\
}

\nakedthanks{$^{\text{\faEnvelopeSquare}}$Contact: Fangcheng Fu <ccchengff@sjtu.edu.cn>}

\maketitle

\begin{abstract}
Agentic sessions driven by Large language models (LLMs) often alternate between model inference and tool use, accumulating long histories across successive rounds.
Serving these sessions efficiently requires reducing attention computation and retaining history key-value (KV) caches to avoid recomputation.
Recently, frontier open-source LLMs adopt sparse attention to reduce computation by selecting only part of the history, while SSDs provide a cheaper alternative to CPU DRAM for storing KV caches.
However, sparse KV selection depends on the ad hoc intermediate values during model inference, so it forces SSD reads to lie on the inference critical path.
These reads are further slowed by fragmented accesses and read-write interference in SSDs.

To address these challenges, we present \system, an agentic serving framework for sparse attention LLMs with SSD-centric KV storage.
\system focuses on append prefill, which processes each round's newly added inputs and accounts for most history KV loading.
To move SSD reads out of the critical path, \system runs the model's own KV selection module on earlier intermediate values, predicting KV demand without additional training.
The predicted reads overlap with model computation, and any prediction misses are fetched before attention executes to preserve model outputs.
To improve SSD efficiency, \system coalesces adjacent reads, packs scattered KV pages into sequential writes on the CPU, and limits background writes while reads are active.
Across three models and three agentic traces, \system outperforms existing works by up to 1.57-3.69$\times$ (1.22-1.85$\times$ on average) in terms of the time to first token latency, while maintaining decode efficiency. 
\end{abstract}

\section{Introduction}
\label{sec:intro}

Large language models (LLMs) have advanced rapidly in reasoning, tool use, and code generation, making autonomous agents one of their most prominent applications. Unlike one-shot generation, an agentic session repeatedly inspects a repository/directory, invokes tools, incorporates their outputs, and revises the solution. Consequently, agentic sessions contain long autonomous loops and increasingly long input contexts~\cite{tracelab}. A recent study reports that agentic tasks consume three orders of magnitude more tokens than chatbots, with input tokens dominating the cost~\cite{agenttokencost}. How to efficiently serve these long-running, multi-round sessions has therefore become an important systems problem.

The first essential property of agentic sessions is their long context. With full attention, processing a sequence of $M$ tokens requires $O(M^2)$ attention computation, which grows prohibitively expensive as an agent accumulates the history context. Sparse attention reduces this cost by allowing each query to attend to only a subset of history tokens, while retaining global or local paths needed for model quality~\cite{nsa}.  For example, top-$k$ sparse attention uses a lightweight indexer to select $k \ll M$ history tokens and reduces the core attention cost to $O(Mk)$.  This approach has moved beyond academic research and has now been used by frontier open-source LLMs, including DeepSeek-V4 and GLM-5 series models~\cite{deepseekv4,glm52}.

The second outstanding property is repeated interaction. Each agent round appends a new user message, tool result, or model output to a much longer session history, and then performs an \emph{append prefill} over the new tokens.  Recomputing the history at every round would waste substantial GPU cycles. Thus, serving systems often retain the history key-value (KV) cache and reload it when the next append prefill query arrives~\cite{lmcache,mooncake}. Due to the limited space of GPU high-bandwidth memory (HBM), current systems commonly offload KV caches to CPU DRAM.  Recent systems further extend this design into a hierarchy spanning GPU HBM, CPU DRAM, and SSDs~\cite{hicache,strata}.

Currently, predominant practices of KV storage are CPU DRAM-centric, treating SSDs or remote storage as lower hierarchies. However, such a design has become costly. Specifically, the ever-increasing context length and user sessions have demanded more and more storage capacity, which drives DRAM to be expensive as demand far exceeds supply. Worse still, the shortage of DRAM is expected to continue given the rapidly growing state of agentic serving. In a snapshot of public server-component prices, a 256-GB DDR5 RAM costs \$8,500--\$14,350, whereas a 30.72-TB enterprise NVMe SSD costs \$14,500--\$16,000~\cite{price_hynix_256g_567n,price_hynix_256g_563n, price_supermicro_256g,price_micron6500_30t,price_solidigm_p5336_30t}. The resulting per-GB cost differs by roughly two orders of magnitude.  This gap has motivated SSD-centric KV storage like Tutti~\cite{tutti}, which leverages GPU-SSD links for KV cache transmission, attempting to approach DRAM-centric serving performance while offering much larger capacity. Consequently, an SSD-centric design that minimizes reliance on CPU DRAM is appealing for cost-efficient agentic serving.

However, it is non-trivial to combine sparse attention with SSD-centric KV storage to serve agentic workloads efficiently. In particular, this work focuses on two essential challenges.

First, \textit{sparse attention makes KV demand non-deterministic to the storage layer and shortens the opportunity to overlap KV loading}. Specifically, since SSD bandwidth is far below HBM bandwidth, SSD-centric systems prefer initiating KV loads early and overlapping them with model computation~\cite{lmcache,strata,tutti}. Dense attention makes this prefetch straightforward as all history KV caches will be consumed. In contrast, sparse attention determines the useful tokens only after its indexer processes the current query for KV selection. Waiting for the exact KV selection places SSD reads on the inference critical path. As we will show in \S\ref{sec:expr}, such SSD reads may occupy over 30\% of the time cost of append prefill. Sparse attention therefore changes not only how many KV caches are read, but also when the serving system can know what to read.

Existing systems do not resolve this mismatch.  Mainstream serving frameworks (e.g., vLLM and SGLang) load the full KV history to HBM so every token remains selectable~\cite{lmcache,hisparse}. This approach maximizes compute--I/O overlap but causes redundant loading, wasting bandwidth and HBM capacity. Predictive approaches instead speculate which tokens will be selected and prefetch only those KV caches~\cite{infinigen,dualdecoder,echo,solidattention,freekv}. Nevertheless, most such designs target the LLM autoregressive decoding phase by exploiting selection similarity between adjacent decode steps, yet our empirical study in \S\ref{sec:obs} reveals agentic workloads exhibit a different bottleneck: nearly all KV reads occur during append prefill, while decode mostly reuses history tokens loaded by append prefill or accesses newly generated tokens in the current round. The few designs that cover prefill~\cite{infinigen,echo} either require offline model calibration or expose only a short overlap window that is insufficient for SSD latency (detailed in \S\ref{sec:obs}).

The second challenge is \textit{the lack of SSD I/O optimization tailored for serving sparse attention LLMs under the agentic scenario}. Specifically, we consider two types of I/O issues, which are the non-contiguous I/O and the read-write interference. For one thing, it is common knowledge that SSDs are inefficient for small random accesses, yet serving sparse attention LLMs naturally produces such unfavorable accesses in both directions. On the read path, only the sparsely selected KV will be loaded, resulting in scattered SSD reads. On the write path, LLM serving maintains newly generated KV caches in non-contiguous GPU pages due to the widely adopted paged attention technique~\cite{pagedattention}, making offloading similarly fragmented. For another, SSDs suffer from the interference of reads and writes as they compete for device-internal resources~\cite{tutti}. Such interference has a larger impact on SSD reads compared to writes, deteriorating read bandwidth by nearly 60\% when concurrent SSD writes exist. Nevertheless, agentic workloads exhibit significant read amplification due to the multi-round nature, which drives read volume to be tens of times greater than write volume (detailed in \S\ref{sec:obs}). Thus, the read-write interference is a particularly pressing problem for agentic workloads. Unfortunately, these I/O issues are under-explored in current systems.

In this work, we present \system\footnote{In Roman myth, \system is usually depicted as having two faces, one looking backward to the past and the other looking forward to the future.}, an agentic serving framework for sparse attention LLMs under the SSD-centric KV storage paradigm. \system addresses the uncertainty and limited overlap opportunity introduced by sparse KV selection, and the I/O issues related to SSD performance, keeping most KV loading off the critical path to make SSD-centric KV storage practical for long-running agent sessions. 

First, \system predicts sparse-attention KV selections early to create an overlap window for SSD loading. During append prefill, it reuses the LLM's own indexer as a training-free predictor by feeding the indexer intermediate states available before its normal execution point. \system then prefetches the predicted KV caches (i.e., predicted reads) while the main model continues computing. Once the model produces the exact KV selection, \system fetches only the KV caches missed by the prediction (i.e., corrective reads), ensuring lossless model outputs. Moreover, as the prediction itself introduces additional computation, \system further manages to accelerate it. Particularly, \system partitions the history context across the participating GPUs and executes the prediction individually. This design has negligible impact on prediction accuracy, but prevents prediction from delaying the inference pipeline.

Second, \system features a series of SSD I/O optimizations tailored for our targeted scenario. To address the non-contiguous I/O, \system coalesces adjacent selected blocks through \textit{best-effort read packing} to trigger contiguous GPU-SSD reads whenever possible, and uses \textit{CPU-assisted write packing} to transform scattered GPU pages into sequential CPU-SSD writes with the help of CPU, avoiding competing with model execution. To address the read-write interference, we apply a \textit{phase-aware write capping} to shelter read performance for agentic workloads.

We implement \system and evaluate it using multiple agentic workload traces and long-context sparse-attention models. The results show that \system limits SSD I/O exposed on the critical path to less than 6.5\% of the append prefill latency, thereby reducing time-to-first-token (TTFT) and outperforming state-of-the-art baselines by up to 1.57-3.69$\times$ (1.22-1.85$\times$ on average).

\section{Preliminaries}
\label{sec:preliminaries}

This section briefly walks through the preliminaries related to our work. Frequently used notations are listed in Table~\ref{tab:notation}.

\begin{table}[t]
\centering
\small
\caption{Frequently used notations. Note that we list notations at token granularity for clarity. The same notation applies when one entry represents a token region.}
\label{tab:notation}
\begin{tabular}{ll}
\toprule
\textbf{Symbol} & \textbf{Description} \\
\midrule
$\intset{n}$ & The set of indices $\{1, 2, \cdots, n\}$. \\
$P$ & Number of tokens in the cached history. \\
$L$ & Number of tokens in the append prefill query. \\
$\mathbf{h}^{(\ell)}_t$ & Input hidden states of token $t$ at layer $\ell$. \\
$\mathbf{z}^{(\ell)}_t$ & Indexer key of token $t$ at layer $\ell$. \\
$\mathbf{k}^{(\ell)}_t,\mathbf{v}^{(\ell)}_t$ & Key and value caches of token $t$ at layer $\ell$. \\
$\mathrm{Idx}^{(\ell)}(\cdot, \cdot; k)$ & Indexer's function at layer $\ell$. \\
$\mathcal{S}^{(\ell)}_t$ & History tokens selected for token $t$ at layer $\ell$. \\
\bottomrule
\end{tabular}
\end{table}

\subsection{LLM Inference and Serving}
\label{sec:prelim_llm_inference_and_serving}

LLM inference consists of two phases. During the \textit{prefill} phase, the model processes all input tokens in one step, computes their KV caches, and generates one output token. The model then enters the auto-regressive \textit{decode} phase, which is executed for several steps, with each step processing the last generated token and generating one output token. In a multi-round session, the KV caches of the existing history can be reused, so only the newly appended tokens are processed. This operation is referred to as \textit{append prefill}. Without loss of generality, we treat the initial prefill of each agentic session as an append prefill with zero history length.

The two phases have different workload characteristics. Prefill is typically compute-bound, especially for long context requests. Consequently, a single long append prefill can already saturate the GPU. In contrast, decode is typically memory-bound, as it repeatedly reads model parameters and KV caches while performing relatively little computation. Thus, modern LLM systems commonly apply the continuous batching technique~\cite{orca_osdi22} to batch multiple requests during decode, while executing each prefill individually.

\begin{figure*}[!t]
\centering
\includegraphics[width=\linewidth]{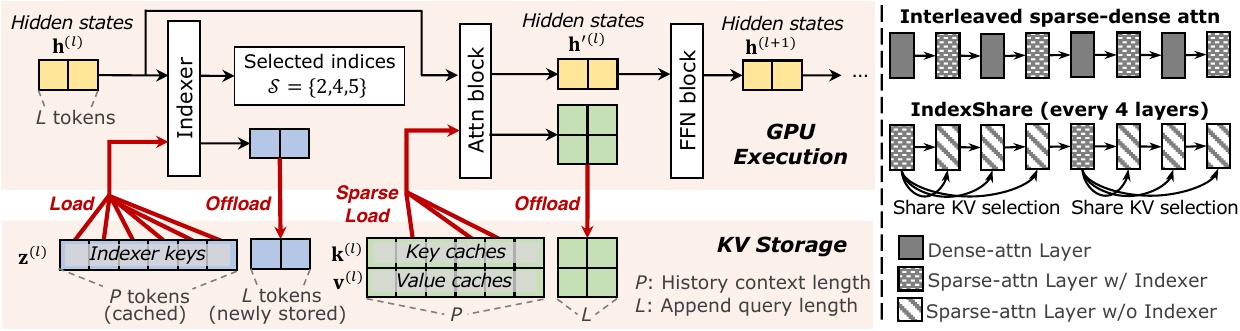}
\caption{Left: Illustration of the sparse-attention Transformer architecture at layer $\ell$. We omit the details (e.g., normalization and residual connection) in attention block (Attn) and feed-forward network block (FFN). Right: Illustration of the interleaved sparse-dense attention design and the IndexShare design that are used in frontier open-source LLMs.}
\label{fig:indexer_attn_ffn}
\end{figure*}

\subsection{Sparse Attention}
\label{sec:prelim_sparse_attention}

Modern LLMs are mainly based on the Transformer architecture. In a nutshell, each Transformer layer comprises an attention block and a position-wise feed-forward network (FFN) block~\cite{vaswani2017attention}, as shown in Figure~\ref{fig:indexer_attn_ffn} (left). For clarity, we denote $\mathbf{h}^{(\ell)}_t, \mathbf{h}^{\prime(\ell)}_t$ as the hidden states for token $t$ at layer $\ell$, where the former is fed into the attention block and the latter is fed into the FFN block. 

As the context grows, the attention mechanism becomes the primary bottleneck due to its quadratic complexity. Recent frontier LLMs use sparse attention to reduce the cost of processing long contexts~\cite{deepseekv32,glm5,deepseekv4,glm52}. Compared to dense attention, sparse attention leverages a lightweight \emph{indexer} to score the relevance between tokens, and then selects a fraction of tokens for the attention mechanism.\footnote{In DeepSeek-V4, sparse selection and attention are performed at the entry granularity, where each entry represents a token region (e.g., 128 consecutive tokens). By default, we describe the mechanism at the token granularity, while our work is implemented at the entry granularity for DeepSeek-V4.}

\mysubsubsection{Prefill phase}
During an append prefill, each sparse attention layer first computes the selection for every appended token. The serving system combines these selections and loads the selected history KV caches into GPU HBM before executing core attention. To be formal, let $P$ be the cached history length and $L$ the appended query length. For each appended token at position $t \in \intset{L}$, the indexer computes its relevance scores to all preceding tokens and selects those with top-$k$ scores, i.e.,
\begin{equation}
\label{eq:indexer_topk}
\mathcal{S}_t^{(\ell)} = \mathrm{Idx}^{(\ell)}\left(\mathbf{h}_t^{(\ell)}, \left\{\mathbf{z}_s^{(\ell)}\right\}_{s\in\intset{P}}; k\right),
\end{equation}
where $\mathcal{S}_t^{(\ell)}$ is the set of selected token indices for token $t$, $\mathrm{Idx}^{(\ell)}(\cdot, \cdot; k)$ represents the indexer's function, $\mathbf{z}_s^{(\ell)}$ is the indexer key for history token $s$.
Indexer keys are per-token states rather than model parameters, and are stored along with KV caches. Before computing Eq.~\eqref{eq:indexer_topk}, the indexer keys for all history tokens must be loaded from the KV storage. Besides, the indexer keys for appended tokens are also derived from the hidden state $\mathbf{h}^{(\ell)}_t$ using the indexer's parameters so that the token selection considers not only the history context but also the preceding tokens within the appended query. We omit such details here and refer interested readers to related technical reports~\cite{deepseekv32,deepseekv4}.

Subsequently, the per-token selections are combined into the union of selection, i.e., $\mathcal{S}^{(\ell)} = \bigcup_{t\in\intset{L}}\mathcal{S}_t^{(\ell)}$, which indicates the indices of all selected tokens for the current layer's sparse attention. Before the attention computation, the KV caches of selected tokens $\{(\mathbf{k}_s^{(\ell)}, \mathbf{v}_s^{(\ell)})\}_{s\in\mathcal{S}^{(\ell)}}$ must be loaded to GPU HBM. The attention block produces intermediate hidden states $\mathbf{h}_t^{\prime(\ell)}$, which will be fed into the FFN block to compute the output hidden states of layer $\ell$, namely $\mathbf{h}_t^{(\ell+1)}$.

In addition, the attention block also generates KV caches $(\mathbf{k}_t^{(\ell)}, \mathbf{v}_t^{(\ell)})$ for each appended tokens $t\in\intset{L}$. After append prefill finishes, the loaded KV caches for selected history tokens and the newly generated KV caches for appended tokens remain in GPU HBM for the ensuing decode phase.

\mysubsubsection{Decode phase}
During the auto-regressive decoding phase, each decode step performs sparse KV selection for the token processed at that step. The routine is similar to an append prefill with a single appended token. If the selection contains a history token whose KV cache is not resident in GPU HBM, the serving system loads that KV cache before executing attention. Otherwise, there will be no KV cache loading. Each newly generated KV cache also remains in GPU HBM for subsequent decode steps.

\mysubsubsection{Uses in frontier open-source LLMs}
Frontier LLMs organize sparse-attention layers in different ways, as illustrated in Figure~\ref{fig:indexer_attn_ffn} (right). DeepSeek-V4 series models interleave sparse-attention layers with dense-attention layers~\cite{deepseekv4}. Each sparse-attention layer performs its own indexer for KV selection, while two consecutive selections are separated by a dense-attention layer. GLM-5 series models instead adopt the IndexShare design: every four consecutive sparse-attention layers form a group, in which only the first layer executes an indexer and the remaining three layers reuse its selected token indices for their respective KV caches~\cite{glm52}. As we will show in \S\ref{sec:method_predictor}, these two designs result in different KV selection prediction performance, which our work takes into account.

\subsection{KV Storage}
\label{sec:prelim_kv_storage}

LLM serving systems retain the KV caches of processed tokens so that subsequent requests can reuse the cached history instead of recomputing it. Because GPU HBM cannot hold the KV caches of many long-running sessions, serving systems commonly offload inactive KV caches to CPU DRAM or SSD and reload them on demand~\cite{lmcache,mooncake,strata,tutti}. Sparse attention LLMs additionally retain the indexer keys described above. Both indexer keys and KV caches are persistent session states. However, the loading of indexer keys is deterministic, so it can be prefetched to overlap the loading. In contrast, only the KV caches of selected history tokens are required, making it infeasible to exactly prefetch those KV caches before the indexer module finishes.

\mysubsubsection{GPU-initiated SSD storage} 
To enhance I/O performance, it is desirable to transfer data directly between GPU HBM and NVMe SSDs, bypassing the CPU. Notably, GPU Direct Storage (GDS) enables peer-to-peer direct memory access (DMA) between HBM and SSD, thereby avoiding a staging copy through CPU DRAM~\cite{gds}. However, GDS still relies on the CPU to initiate each I/O request. In response to this, GPU-initiated SSD storage systems such as GeminiFS and Tutti expose NVMe queues to the GPU, allowing GPU kernels to submit I/O commands and observe their completion without per-request CPU intervention~\cite{geminifs,tutti}. The CPU remains responsible for setup and coarse-grained metadata management outside the data path. 

In particular, NVMe communicates with GPU kernels through queue pairs, each pair consisting of a Submission Queue (SQ) and a Completion Queue (CQ). To issue an I/O, a GPU kernel enqueues a command in the SQ that describes the I/O request (e.g., the operation, source and destination addresses, and transfer length). The GPU then rings the SQ's doorbell to notify the NVMe controller. The controller fetches the pending command(s), accesses the SSD, and transfers the data directly between SSD and HBM through peer-to-peer DMA. After the transfer finishes, the controller posts a completion singal to the CQ, which the GPU polls to determine when the requested I/O is done. 
This queue-based solution allows GPU kernels to submit asynchronous I/O requests and overlap storage access with GPU computation.

\section{Observations and Motivation}
\label{sec:obs}

Agentic serving differs from conventional multi-turn chat in both its input and output lengths. We first characterize this difference and then present two observations about its impact on SSD-backed sparse-attention serving, which motivate the design of \system.

\mysubsubsection{Length characteristics of agentic workloads}
Conventional multi-turn conversations usually append a short user message and generate a longer model response. Agentic sessions often exhibit the opposite pattern. An appended query can contain the output of a tool, such as a PDF extractor or web searcher, or the result returned by a sub-agent. Such outputs can be much longer than a user message. Meanwhile, an agent decomposes a complex task into multiple steps, and each step often generates only a tool invocation, an intermediate decision, or a short partial result. Agentic workloads therefore commonly combine long inputs, short outputs, and many rounds.

\begin{table}[!t]
\centering
\small
\caption{Token-length distributions in three agentic traces (detailed in \S\ref{sec:expr_setup}), where $P$ is the history context length before an agent round, $L$ is the append query length, and $O$ is the output length in that round.}
\label{tab:agentic_length}
\resizebox{\linewidth}{!}{
\begin{tabular}{lrrr|rrr|rrr}
\toprule
 & \multicolumn{3}{c|}{Trace 1} & \multicolumn{3}{c|}{Trace 2} & \multicolumn{3}{c}{Trace 3} \\
\cmidrule{2-4} \cmidrule{5-7} \cmidrule{8-10}
 & $P$ & $L$ & $O$ & $P$ & $L$ & $O$ & $P$ & $L$ & $O$ \\
\midrule
Avg. & 77K  & 1.1K & 237 & 119K & 1.4K & 365 & 125K & 4.8K & 357 \\
P25  & 71K  & 742  & 142 & 85K  & 727  & 222 & 67K  & 856  & 98  \\
P50  & 77K  & 901  & 228 & 131K & 1.3K & 404 & 70K  & 2.2K & 216 \\
P90  & 85K  & 1.8K & 401 & 154K & 2.7K & 540 & 323K & 12K  & 671 \\
P99  & 88K  & 2.3K & 510 & 155K & 4.0K & 590 & 375K & 22K  & 1.1K \\
\bottomrule
\end{tabular}
}
\end{table}

\begin{figure}[!t]
\centering
\includegraphics[width=\linewidth]{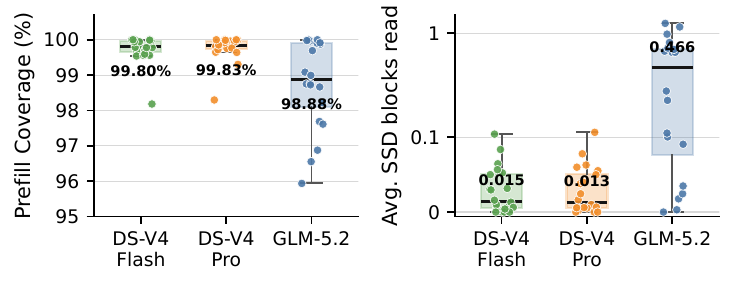}
\caption{Comparison of KV loading between append prefill and decode over 20 agentic rounds, evaluated on DeepSeek-V4-Flash, DeepSeek-V4-Pro, and GLM-5.2. Left: The coverage of history KV caches that are required by decode phase but loaded in append prefill. Right: Average numbers of SSD blocks loaded per sparse attention during decoding.}
\label{fig:obs1}
\end{figure}

Table~\ref{tab:agentic_length} confirms this pattern in three agentic traces. Their median history lengths reach 77K, 131K, and 70K tokens, respectively. The append query length and model output length are relatively short compared to the history. However, the former is still 3-20$\times$ of the latter in both median and P99 values.
These distributions align with recent characterizations of coding-agent workloads, which report long contexts and short outputs~\cite{tracelab}.

This length asymmetry has two implications for KV storage. First, a thousand-token append query can require a large set of history KV caches during append prefill, whereas the short decode can reuse much of this set. Second, repeated agent rounds can load the history KV caches multiple times over the lifetime of a session. We next quantify these two effects.

\mysubsubsection{Observation 1: Append prefill dominates KV loading}
Both append prefill and decoding perform sparse KV selection and may load history KV caches, as described in \S\ref{sec:prelim_kv_storage}. To identify the dominant phase, we record the history KV volume transferred from SSD to GPU HBM during representative agent rounds. We separately aggregate the transfers triggered by append prefill and all decode steps in the same round. Our measurement counts only actual history KV transfers and excludes KV caches already resident in GPU HBM and KV caches newly generated in the current round. 

We randomly sample 20 agentic rounds from the traces and record the sparse KV loading during append prefill and the subsequent decoding steps. As shown in Figure~\ref{fig:obs1}, across the three models, roughly 99\% of the history KV caches required by the decode phase are already loaded in append prefill. Even though decode often spans hundreds of steps, most decode steps trigger no KV loading at all. Instead, they reuse the history KV caches loaded during append prefill and the KV caches generated earlier in the same round.

The long-input, short-output pattern explains this concentration. An append prefill computes sparse selections for many appended tokens, and their union brings a broad history working set into GPU HBM. This working set remains resident during the ensuing decode phase. Subsequent decode steps therefore tend to select KV caches already in HBM or KV caches newly generated in the current round, leaving little additional data to load from SSD.

This observation shifts the optimization target from decoding to append prefill. Most existing KV prefetchers focus on decoding and predict future selections from their similarity across adjacent decode steps~\cite{dualdecoder,freekv,solidattention,hisparse}, which cannot be leveraged in append prefill. 
There are few predictive prefetchers for append prefill, yet they still fail to provide both lossless execution and a sufficiently long SSD overlap window. InfiniGen~\cite{infinigen} performs model-specific offline calibration so that attention scores can be estimated for KV selection. However, it requires specific model tuning and cannot guarantee the output to be lossless. 
ECHO~\cite{echo} partitions prefill queries into segments, performs per-segment KV selection prediction, and overlaps the current segment's prefetching (from CPU DRAM) with the next segment's prediction. However, this requires the indexer computation could overlap sparse KV reads, which is infeasible for SSD storage since sparse KV reads take longer time than indexer (detailed in \S\ref{sec:expr_e2e}).

\begin{figure}[!t]
\centering
\includegraphics[width=\linewidth]{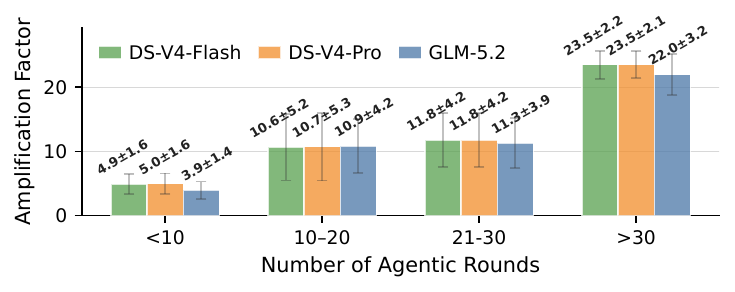}
\caption{The read amplification factors (i.e., cumulative read volume divided by cumulative write volume) under different agentic rounds.}
\label{fig:obs2}
\end{figure}

\mysubsubsection{Observation 2: Agentic sessions amplify KV reads}
The multi-round nature of agentic sessions creates \emph{KV read amplification}. A token's KV cache is written to SSD once after it is generated and evicted from GPU HBM. However, that KV cache can be selected and reloaded by multiple later append prefills as the agent accumulates tool results and intermediate decisions. Long appended queries further increase the history KV demand in each round. Consequently, the cumulative amount of KV data read from SSD can exceed the amount written to SSD over a complete session.

Sparse attention reduces the history KV cache loaded in each round, but does not eliminate this amplification. It limits each appended token to a subset of the history, yet the same history KV cache can still appear in the selected subsets of many later rounds. Figure~\ref{fig:obs2} compares the cumulative KV reads and writes of sparse attention from real agentic traces. It can be seen that the total read volume remains 3.9-23.5 times of the write volume despite sparse selection. And the amplification enlarges as the number of agentic rounds increases.

KV read amplification makes SSD read efficiency particularly important compared to SSD write efficiency. Moreover, full-duplex data paths do not isolate reads and writes inside an SSD: concurrent requests still compete for device-internal resources and degrade I/O efficiency~\cite{tutti}. Sparse attention further scatters reads across the selected history, while paged KV allocation scatters writes across non-contiguous GPU pages~\cite{pagedattention}. These properties make the repeated reads induced by agentic sessions difficult to execute efficiently.

Together, our observations raise two important questions. First, how to hide the sparse KV loading during append prefill rather than decode, as it dominates the I/O exposed on the critical path. Second, how to emphasize read efficiency when optimizing SSD I/O for our target scenario owing to the read amplification of agentic workloads. These also motivate the core designs of \system introduced in \S\ref{sec:method}.

\section{\system}
\label{sec:method}

\subsection{Overview}
\label{sec:method_overview}

\begin{figure}[!t]
\centering
\includegraphics[width=\linewidth]{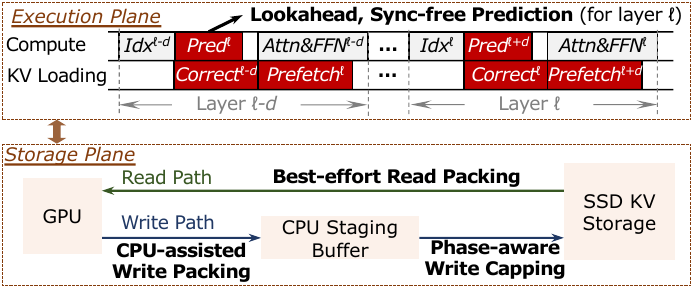}
\caption{Overview of \system.}
\label{fig:overview}
\end{figure}

Figure~\ref{fig:overview} depicts the overview of \system, which mainly consists of the execution plane and the storage plane.

The core of the execution plane is our KV selection prediction (\S\ref{sec:method_predictor}). When an append prefill request arrives, \system triggers the prefetching for deterministic KV loading (e.g., indexer keys and KV caches for dense attention layers), and adopts \textit{training-free lookahead prediction} to facilitate prefetching of KV caches of sparse attention. Specifically, the KV selection of a target layer is predicted a few layers ahead, guided by a \textit{model-aligned prediction distance}. In addition, to reduce prediction overhead, we develop \textit{synchronization-free parallel prediction}, which parallelizes prediction while removing cross-GPU synchronization. Based on the prediction results, GPUs trigger predictive reads to prefetch the sparsely selected KV caches. When execution reaches the target layer, its indexer computes the exact KV selection. \system issues corrective reads for missing KV caches before executing sparse attention in order to preserve lossless model outputs. After append prefill, the request enters decode. 

The storage plane provisions SSD I/O optimization techniques tailored for both read and write paths by taking the characteristics of sparse LLM agentic workloads into account (\S\ref{sec:method_io_optim}). For predictive and corrective reads, which involve non-contiguous GPU-SSD reads, \system applies \textit{best-effort read packing} to merge adjacent blocks whenever possible. To offload the newly generated KV caches, \system leverages a \textit{CPU-assisted write packing} approach, which allows for contiguous CPU-SSD writes without competing model execution on the GPU. Moreover, to maintain read performance for agentic workloads, \system employs a simple-yet effective \textit{phase-aware write capping}, which limits background write bandwidth when reads are in flight.

\subsection{KV Selection Prediction}
\label{sec:method_predictor}

\mysubsubsection{Training-free lookahead prediction}
As mentioned in \S\ref{sec:obs}, prefetching from SSD requires a sufficient overlap window. \system achieves this by invoking the target layer's own indexer before its normal execution point. Denote the prediction distance as $d$, i.e., predicting KV selection for the sparse attention of layer $\ell$ at an earlier layer $\ell-d$. Let $\mathbf{h}^{(\ell-d)}_t$ denote the per-token hidden state at layer $\ell-d$.
\system feeds this earlier hidden state to the target layer's indexer while retaining the target layer's indexer keys $\mathbf{z}^{(\ell)}_s$. Formally, our lookahead prediction can be written as
\begin{equation}
\label{eq:predicted_kv_selection}
\widehat{\mathcal{S}}^{(\ell)}_t = \mathrm{Idx}^{(\ell)}\left(\mathbf{h}_t^{(\ell-d)}, \left\{\mathbf{z}_s^{(\ell)}\right\}_{s=1}^{P}; k\right)
\end{equation}
For each target layer, \system prefetches the union $\widehat{\mathcal{S}}^{(\ell)} = \bigcup_{t\in\intset{L}}\widehat{\mathcal{S}}^{(\ell)}_t$ while the normal model execution proceeds. When layer $\ell$ later computes the exact KV selection $\mathcal{S}^{(\ell)}$, \system issues corrective reads for selected entries absent from the predicted union, i.e., $\mathcal{S}^{(\ell)} \setminus \widehat{\mathcal{S}}^{(\ell)}$. Thus, the amount of SSD I/O exposed on the critical path is reduced, while the KV caches consumed by sparse attention remain unaltered, ensuring lossless model output. The lookahead prediction introduces no auxiliary model or parameters, requires no training or offline calibration, and turns the distance $d$ into an overlap window for predictive reads.

\begin{figure}[!t]
\centering
\includegraphics[width=\linewidth]{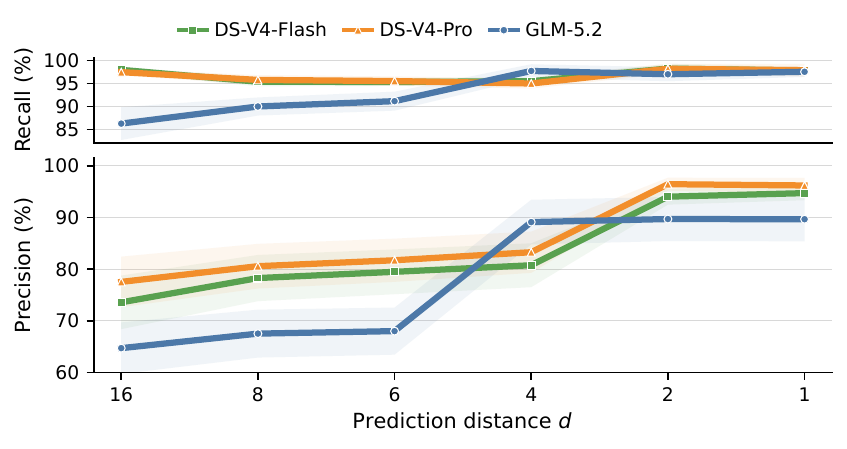}
\caption{Prediction quality w.r.t. prediction distance ($d$). A higher recall indicates fewer corrective reads, and a higher precision indicates less redundancy in predictive reads.}
\label{fig:prediction_distance}
\end{figure}

\mysubsubsection{Model-aligned prediction distance}
We empirically evaluate how the prediction distance $d$ affects the precision and recall of the predicted KV selection. Recall measures the fraction of the exact KV union covered by predictive reads (i.e., $\lvert \widehat{\mathcal{S}}^{(\ell)} \cap \mathcal{S}^{(\ell)} \rvert / \lvert \mathcal{S}^{(\ell)} \rvert$) and thus determines the amount of corrective reads, whereas precision measures the fraction of prefetched KV caches that are useful (i.e., $\lvert \widehat{\mathcal{S}}^{(\ell)} \cap \mathcal{S}^{(\ell)} \rvert / \lvert \widehat{\mathcal{S}}^{(\ell)} \rvert$) and thus reflects the redundancy in predictive reads. Figure~\ref{fig:prediction_distance} reports the results across three models. Overall, a smaller $d$ substantially improves precision, while recall is comparatively less sensitive and remains high, especially for DeepSeek-V4. Moreover, both metrics eventually saturate at short prediction distances, exhibiting diminishing returns.

It is noteworthy that the high recall at a large $d$ does not necessarily indicate sound prediction. Although the indexer makes top-$k$ selection for each of the $L$ appended tokens, because semantically related tokens tend to have overlapping selection, the union of selection is typically smaller than $L \times k$. At a large $d$, the top-$k$ prediction for each appended token becomes more random, which leads to less overlapping in the selection, inflating the predicted union. This larger union can cover many exact tokens through broader coverage and thereby preserve recall, but it also redundantly fetches many KV caches that are never used, as reflected by the low precision. 
Consequently, we seek the prediction distance that achieves high precision and recall.

Figure~\ref{fig:prediction_distance} further reveals that the point of diminishing returns aligns with model designs. For both DeepSeek-V4 models, reducing $d$ from 2 to 1 changes the two metrics by less than $0.7\%$. This saturation distance aligns with their interleaved sparse--dense attention architecture (i.e., consecutive sparse attention layers are separated by two layers). For GLM-5.2, the saturation distance is 4, aligning with its IndexShare design that shares top-$k$ selections across four consecutive layers.

Motivated as such, we set $d=2$ for DeepSeek-V4 and $d=4$ for GLM-5.2, provisioning sound prediction quality while allowing predictive reads to be issued as early as possible.
By doing so, the recall metric is above 96\% for all three models, indicating that there are only 4\% of the sparse KV loading remaining on the critical path. Meanwhile, we keep the redundancy in predictive reads below 5\% for DeepSeek-V4 and 10\% for GLM-5.2.

\begin{figure}[!t]
\centering
\includegraphics[width=0.9\linewidth]{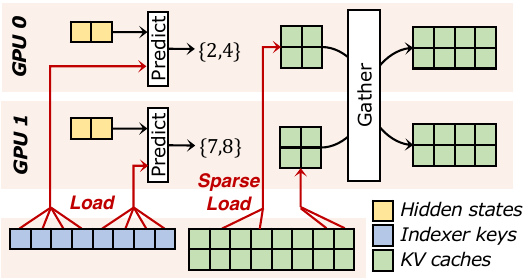}
\caption{Illustration of our synchronization-free prediction.}
\label{fig:sync_free_predict}
\end{figure}

\begin{figure}[!t]
\centering
\includegraphics[width=\linewidth]{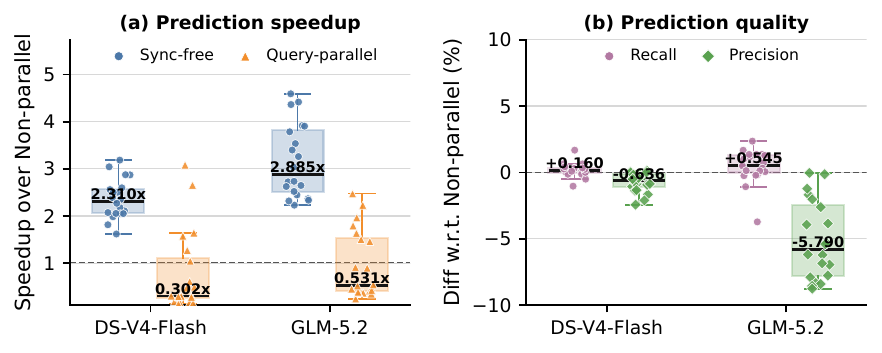}
\caption{Assessment of synchronization-free prediction on 8 GPUs. (a) Prediction speedup over non-parallel prediction of synchronization-free and query-parallel prediction methods. (b) The differences in prediction quality of synchronization-free prediction compared to non-parallel prediction.}
\label{fig:parallel_prediction_results}
\end{figure}

\mysubsubsection{Efficient synchronization-free parallel prediction}
Although our lookahead prediction creates the overlap window for prefetching, it incurs non-negligible computation overhead. To be specific, it requires to compute the relevance scores for $L$ appended tokens against their preceding tokens, resulting in a computation complexity of $O(PL)$ given that $P>L$. Such prediction overhead would harm the overall performance. Considering that frontier LLMs are commonly deployed on multiple GPUs due to their model scale, we manage to distribute prediction across these GPUs so that each GPU performs approximately $1/N$ of the computation, where $N$ is the number of GPUs to deploy one model replica.

\begin{figure*}
\begin{minipage}{0.32\textwidth}
\includegraphics[width=\linewidth]{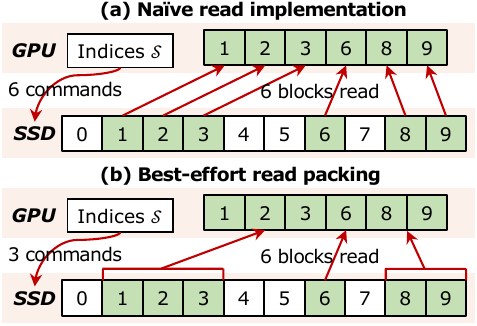}
\caption{Illustration of read packing.}
\label{fig:read_packing}
\end{minipage}
\begin{minipage}{0.01\textwidth}
$ $
\end{minipage}
\begin{minipage}{0.66\textwidth}
\includegraphics[width=\linewidth]{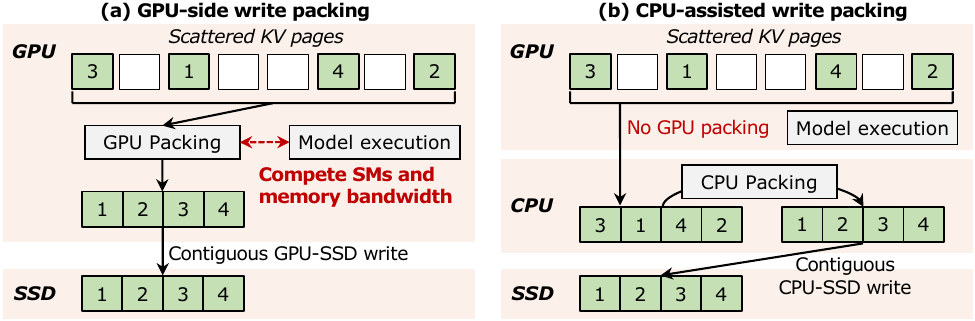}
\caption{Illustration of GPU-side write packing and CPU-assisted write packing.}
\label{fig:write_packing}
\end{minipage}
\end{figure*}

However, how to efficiently parallelize the prediction necessitates thoughtful deliberation. Below we analyze two natural partitioning dimensions. 
\begin{itemize}[leftmargin=*]
\item \emph{Query-parallel partitioning} distributes the hidden states $\mathbf{h}^{(\ell-d)}$ for the $L$ appended query tokens across GPUs. Each GPU predicts the selection for its assigned query tokens, after which the GPUs must merge and deduplicate their selected indices before prefetching to avoid redundant reads. 
\item \emph{Indexer-key-parallel partitioning} shards the indexer keys $\mathbf{z}^{(\ell)}$ for the $P$ cached history tokens. Each GPU computes the relevance scores for all appended tokens against one history shard and produces a local top-$k$ list for each appended token. Then, the GPUs gather the local lists (along with the local top-$k$ relevance scores) to obtain the exact global top-$k$ list for each appended token, and finally obtain the union of selected tokens.
\end{itemize}
As a result, although both partitioning approaches can reduce the computation on each GPU, they incur a cross-GPU synchronization point between the prediction and SSD prefetching. Although the communication volume among the GPUs is not high, synchronization still prevents us from speeding up prediction via parallelization.

Fortunately, we find that our lookahead prediction can be synchronization-free under index-key-parallel partitioning. This is because speculative prefetching does not require the exact global top-$k$. Motivated by this, we develop a synchronization-free parallel prediction by turning the global top-$k$ computation into $N$ individual top-$k/N$ processes across the GPUs, as shown in Figure~\ref{fig:sync_free_predict}. 

Formally, denote $\mathcal{C}_n$ as the history shard assigned to GPU $n$. We first let each GPU produce the local top-$k/N$ list for each appended token based on the corresponding history shard, i.e., 
\begin{equation}
\widehat{\mathcal{S}}_{t}^{(\ell, n)} = \mathrm{Idx}^{(\ell)}\left(\mathbf{h}_t^{(\ell-d)}, \left\{\mathbf{z}_s^{(\ell)}\right\}_{s \in \mathcal{C}_n}; k/N\right)
\end{equation}
Then, instead of synchronizing the local results into the global one, GPU $n$ directly computes the union of local selected indices $\widehat{\mathcal{S}}^{(\ell, n)} = \bigcup_{t\in\intset{L}} \widehat{\mathcal{S}}_{t}^{(\ell, n)}$. Subsequently, each GPU immediately triggers the prefetching individually. Since different shards cover disjoint history tokens, there will be no duplicated reads. Finally, all GPUs gather the locally selected states using IPC before the target layer consumes them.

It is undoubted that removing the synchronization cannot ensure exact global top-$k$ selection for each appended token, since a certain shard may contain more than $k/N$ high-scoring history tokens. However, after combining the token selection for all appended tokens, the approximation errors become minor. Our empirical results in Figure~\ref{fig:parallel_prediction_results} show that, the synchronization-free parallel prediction achieves almost the same recall compared to non-parallel prediction, with a small drop in precision. In terms of efficiency, our approach consistently accelerates the prediction, achieving speedup by 2.8$\times$ and 2.3$\times$ on average for GLM-5.2 and DeepSeek-V4-Flash, respectively (results on both DeepSeek-V4 models are similar). In contrast, the query-parallel approach slows down the prediction in many cases due to synchronization. 

Last but not least, our approach further helps reduce the SSD I/O per GPU. As depicted in Figure~\ref{fig:sync_free_predict}, each GPU only needs to load its own history shard and its local selected KV caches, rather than loading the full KV history. Therefore, \system parallelizes both prediction and SSD reads.

\subsection{SSD I/O Optimization}
\label{sec:method_io_optim}

In this section, we introduce how \system addresses the non-contiguous I/O and read-write interference for serving sparse attention LLMs under agentic workloads.

\mysubsubsection{Best-effort read packing} 
As mentioned in \S\ref{sec:prelim_kv_storage}, to initiate an SSD read on the GPU side, the GPU needs to enqueue I/O commands into the NVMe SQ and then rings the doorbell. 
A na\"ive implementation maps every selected SSD block to an individual command, as shown in Figure~\ref{fig:read_packing}(a). Although the GPU can trigger these commands with a single doorbell ring, the SSD still executes many small reads, which underutilize its bandwidth and incur per-command processing overhead.

\system instead converts the KV selection into an address-ordered list of unique SSD blocks and identifies maximal runs of consecutively selected blocks. As illustrated in Figure~\ref{fig:read_packing}(b), \system submits one read command for each run rather than one command for each block. For example, three selected blocks at consecutive addresses are fetched with one request covering all three blocks. This packing is best-effort: \system merges only blocks that are both selected and physically adjacent, without fetching unselected blocks merely to enlarge a request. Consequently, read packing increases the average request size while reducing the number of NVMe commands as well as their pressure on the device queues. When the selected blocks are fully scattered, it naturally falls back to the original block-granular reads without introducing additional I/O.

\mysubsubsection{CPU-assisted write packing} 
Paged attention allocates the KV caches of newly generated tokens to non-contiguous GPU pages~\cite{pagedattention}. Directly offloading these pages to SSD requires many small, scattered writes, which not only underutilize SSD bandwidth but also consume NVMe queue entries and repeatedly incur the fixed latency of an I/O request~\cite{tutti}. A natural alternative is to relayout the pages into a contiguous GPU buffer before issuing a large GPU-to-SSD write, as shown in Figure~\ref{fig:write_packing}(a). However, this relayout requires a GPU kernel that competes with model execution for streaming multiprocessors (SMs) and memory bandwidth, which is undesirable.

\system therefore offloads write packing to the CPU, as illustrated in Figure~\ref{fig:write_packing}(b). When KV offloading occurs, \system first transfers the scattered KV pages asynchronously to a bounded CPU staging buffer. The CPU then relayouts these pages in token order, after which \system writes the resulting contiguous buffer to the SSD with one sequential request. This removes packing work from the GPU and avoids scattered page-granular SSD writes. Moreover, the resulting token-ordered SSD placement creates the contiguous regions that best-effort read packing can exploit in subsequent append prefills.

It is noteworthy that the CPU staging buffer does not rely on a large CPU DRAM space. Our empirical results in \S\ref{sec:expr} show that only 16 GiB of CPU staging buffer is already sufficient for serving large-scale models like DeepSeek-V4-Pro, which has 1.6 trillion parameters. The buffer holds only KV caches currently moving from GPU memory to SSD and is recycled after their writes complete. Therefore, CPU-assisted packing avoids the interference between GPU-side packing and model computation, while retaining the cost-efficiency advantage of SSD-centric KV storage.

\begin{figure}[!t]
\centering
\includegraphics[width=\linewidth]{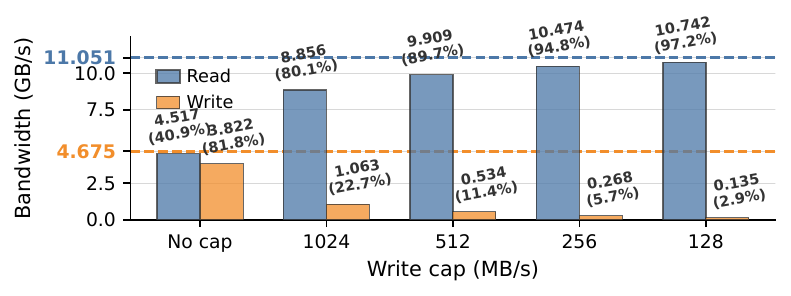}
\caption{The read and write bandwidth under different write caps. The blue and orange lines indicate the read and write bandwidth without interference, respectively. Values in parentheses are the portion w.r.t. bandwidth without interference.}
\label{fig:read_write_interference_results}
\end{figure}

\mysubsubsection{Phase-aware write capping} 
Unfortunately, there exists interference between SSD reads and writes. Concurrent requests compete for device-internal resources, particularly the NVMe internal cache, and reduce both read and write bandwidth~\cite{tutti}. As shown in Figure~\ref{fig:read_write_interference_results}, when read and write interfere with each other, read bandwidth is degraded by nearly 60\%, which is more severely than write bandwidth is. Thus, this interference is especially harmful to agentic serving due to the read amplification described in \S\ref{sec:obs}. Moreover, prefetch reads must be overlapped within the time window, and corrective reads issued after the exact indexer lie directly on the critical path. Write traffic that slows down reads would therefore degrade the append prefill efficiency substantially.

To mitigate this interference, \system applies a simple yet effective phase-aware write limit. Specifically, we observe that by capping the write bandwidth, we can restore read bandwidth concurrent write traffic. Motivated by this, during append prefill, \system caps background writes at 512 MB/s per SSD, preserving about 90\% of the read bandwidth for prefetch and corrective reads. To be specific, we split the write message into chunks of 256MB. Whenever the CPU is about to write a chunk, it probes whether there are inflight SSD reads. If yes, the write cap is enforced, and vice versa. By doing so, \system prioritizes the read efficiency without permanently sacrificing write throughput.

\subsection{Implementation}
\label{sec:method_impl}

\system is an agentic serving framework for sparse attention LLMs with SSD-centric KV storage, implemented with 44K lines of Python and C++/CUDA codes. We integrate Tutti~\cite{tutti} to support GPU-initiated SSD I/O while CPU-SSD writes are implemented using standard POSIX file I/O. \system uses vLLM~\cite{2026vllm_github} as the backend engine for LLM inference and serving, yet other engines like SGLang~\cite{sglang} are also compatible. \system supports frontier sparse attention LLMs including DeepSeek-V4 series and GLM-5 series models~\cite{deepseekv4,glm5}. We perform layer-wise prefetching for deterministic loading, including indexer keys, KV caches for dense attention layers, and the sliding window KV caches used in DeepSeek-V4. For sparsely selected KV caches, we speculatively prefetch them using our lookahead prediction and trigger corrective reads immediately after the indexer computes the exact KV selection. Furthermore, we invoke lookahead selection prediction during corrective reads (as depicted in Figure~\ref{fig:overview}) to hide the prediction overhead, and we leverage GPU copy engines to perform the KV gathering after each GPU loads its own shard, avoiding SM contention with model execution.

\section{Experiments}
\label{sec:expr}

\begin{figure*}[!t]
\centering
\includegraphics[width=1.03\linewidth]{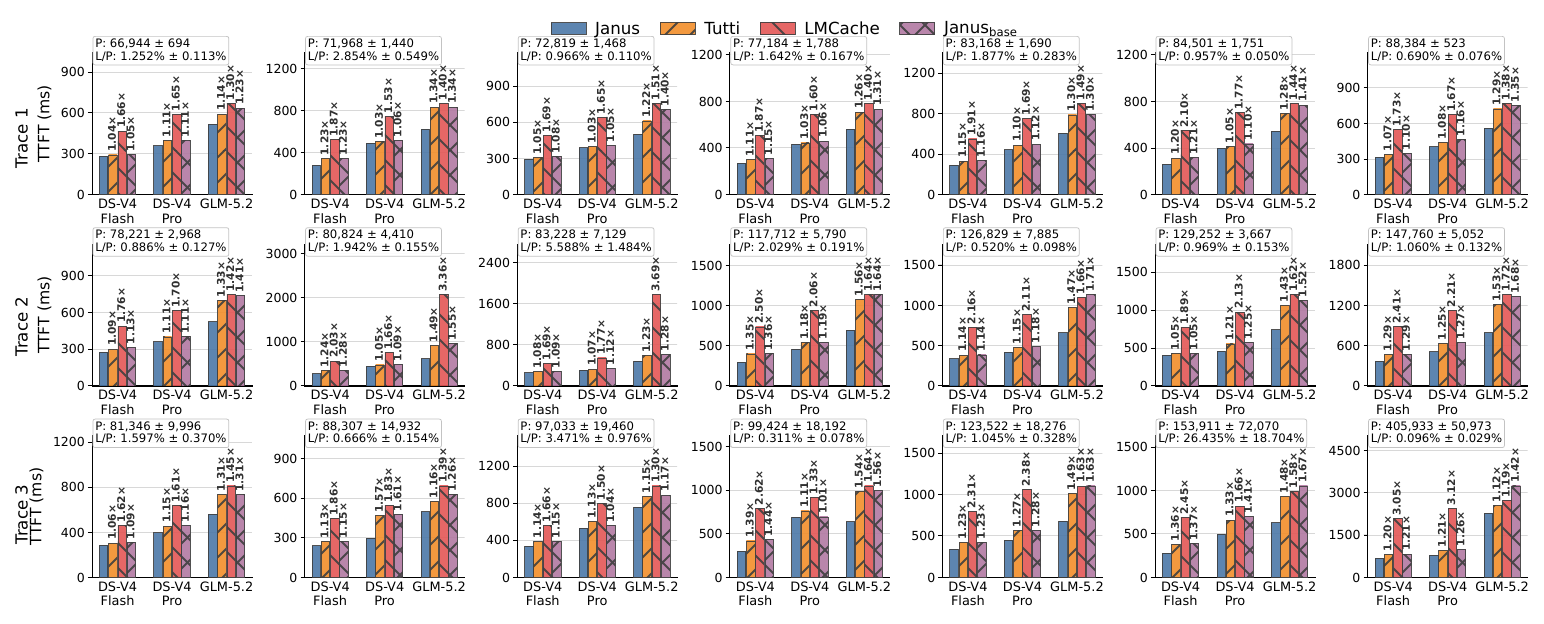}
\caption{End-to-end TTFT for isolated append-prefill requests. For each trace, we cluster requests into seven groups by history length $P$ and append-to-history ratio $L/P$, shown above each panel, and report the mean TTFT using bar plots within each group. The value above each bar is the speedup achieved by \system compared to the corresponding baseline.}
\label{fig:expr_e2e_ttft}
\end{figure*}

\subsection{Experimental Setup}
\label{sec:expr_setup}

\mysubsubsection{Hardware environments}
We conduct experiments on one GPU server equipped with 8 H200 GPUs (each with 141 GB HBM), 1TB CPU DRAM, and 8 KIOXIA CD8P-R 7.68 TB SSDs. The GPUs are interconnected with 900 GB/s NVLink, and the other interconnects use PCIe 5.0.

\mysubsubsection{Models}
Three frontier open-source LLMs are used in our experiments, which are DeepSeek-V4-Flash, DeepSeek-V4-Pro, and GLM-5.2~\cite{deepseekv4,glm52}, containing 284B, 1.6T, and 753B parameters, respectively. Model parameters and KV states are stored in FP8, while indexer keys of DeepSeek-V4 are stored in FP4 following their default option~\cite{deepseekv4}.

\mysubsubsection{Traces}
Our evaluation uses the three agentic traces listed in Table~\ref{tab:agentic_length}. The first two traces are generated\footnote{Existing publicly available traces for agentic workloads (e.g.,~\cite{tracelab,agentx}) contain per-round length information but do not provide raw texts/tokens, which are necessary for computing the sparse KV selection. Thus, we generate traces from benchmarks and we will open-source them.} from SWE-Bench-Pro and Terminal-Bench~\cite{swebench_pro,terminalbench}, while the third one is a proprietary trace collected from our industrial partner. 

\mysubsubsection{Baselines}
We mainly compare \system against Tutti~\cite{tutti} and LMCache~\cite{lmcache}. Tutti is an SSD-centric KV storage for LLM serving, which allows GPU to initiate SSD reads and writes without CPU intervention. LMCache is a widely used KV management framework that supports GPU-CPU-SSD hierarchy, but requires CPU to initiate each I/O request. In addition, since both Tutti and LMCache only support loading the full history KV at once, we further consider a variant of \system, denoted as \systembase. This variant does not adopt prediction, but trigger sparse KV reads only after the indexer finishes KV selection. All competitors use vLLM~\cite{2026vllm_github} as the backend, and we restrict available CPU DRAM capacity to be 16 GB to focus our evaluation on SSD storage. Specifically, we configure LMCache to use SSDs only, as hierarchical storage has much worse performance under the DRAM strict.

\mysubsubsection{Evaluation metrics}
Since we focus on the efficiency of append prefill, we mainly report time to first token (TTFT) to demonstrate how \system accelerates append prefill. We also report time per output token (TPOT) when evaluating \system in serving to show that our work maintains decode efficiency. All experiments are averaged over five runs.

\subsection{End-to-end Comparison}
\label{sec:expr_e2e}

\mysubsubsection{Append prefill efficiency}
\label{sec:expr_ttft}
We first assess the time cost of append prefill. For each trace, we cluster append prefill requests with similar history lengths and append-to-history ratios because both dimensions affect the computation and sparse KV reads. We then compare the mean TTFT of all competitors. 

As shown in Figure~\ref{fig:expr_e2e_ttft}, the speedup achieved by \system varies among clusters because history length and append ratio change both the sparse KV working set and the amount of model computation available for overlap. Nevertheless, \system consistently achieves the lowest TTFT across all length characteristics. Specifically, \system provides up to 1.57$\times$ (1.22$\times$ on average) speedup over Tutti, up to 3.69$\times$ (1.85$\times$ on average) speedup over LMCache, and up to 1.71$\times$ (1.27$\times$ on average) speedup over \systembase.

As LMCache relies on CPU to initiate GPU-SSD reads and writes, it suffers from the highest TTFT in most cases due to the extra framework-side overhead. Tutti and both variants of \system support GPU-initiated SSD I/O to eliminate the related overhead. Tutti fetches all history KV while \systembase loads only the sparsely selected KV. Theoretically speaking, Tutti results in redundant I/O compared with \systembase. However, we observe that \systembase has higher TTFT in many cases. This is not surprising: \systembase leaves all sparse KV reads on the critical path while the redundant loading of Tutti can be partially overlapped with model computation. For \system, although corrective reads still lie on the critical path, by speculatively prefetching the sparsely selected KV, \system successfully hides most SSD I/O with model computation, and therefore outperforms all baselines in terms of TTFT latency.

\begin{figure}[!t]
\centering
\includegraphics[width=\linewidth]{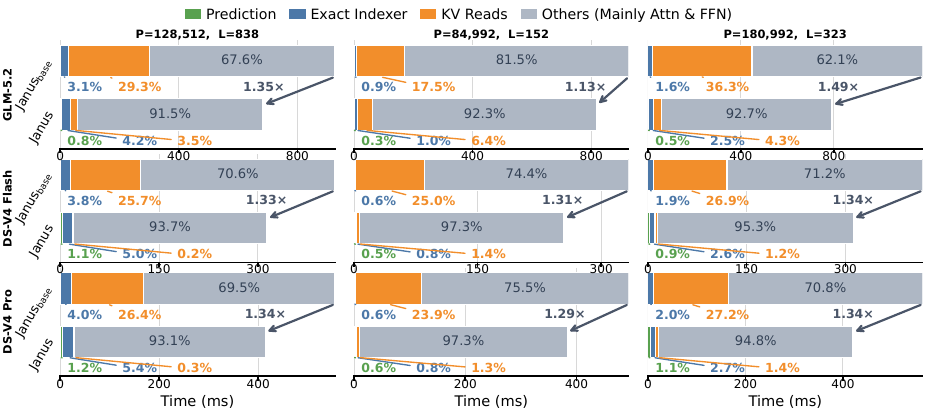}
\caption{Time breakdown for three representative append prefill requests with different history lengths ($P$) and append lengths ($L$). Prefetched KV reads overlap with computation and are omitted from \system's bars; its KV-read segments contain only corrective reads. Green arrows show the speedup over \systembase.}
\label{fig:expr_case_studies}
\end{figure}

\begin{figure}[!t]
\centering
\includegraphics[width=\linewidth]{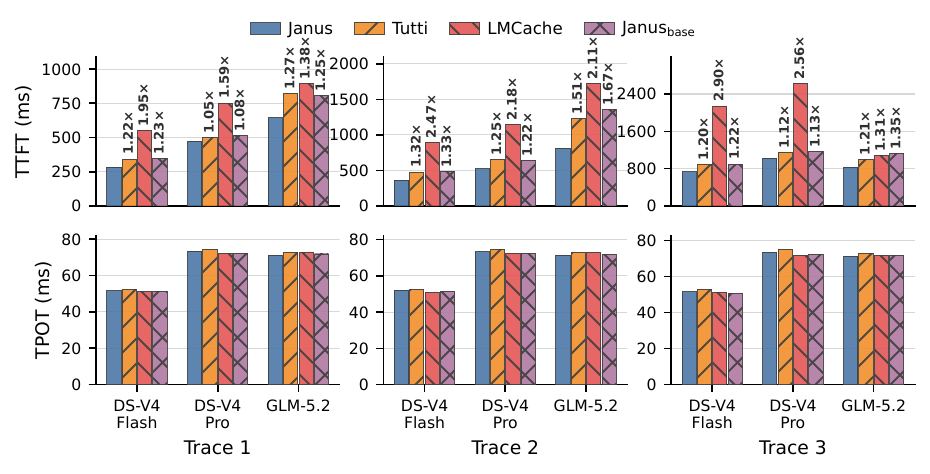}
\caption{P95 TTFT and TPOT under end-to-end serving. The value above each bar is the speedup achieved by \system compared to the corresponding baseline.}
\label{fig:expr_e2e_serving}
\end{figure}

\mysubsubsection{Case studies}
To understand where the TTFT reduction comes from, Figure~\ref{fig:expr_case_studies} decomposes the critical path of three representative append prefill requests across the three models. \systembase first executes the exact indexer, then reads the selected sparse KV caches before attention can proceed. These reads account for 17.5--36.3\% of the time cost of append prefill. Thus, even GPU-initiated reads expose a substantial non-compute interval when they start only after the exact selection becomes available.

\system converts most of this interval into overlapped I/O. Its prediction runs before the target layer, allowing predicted KV reads to proceed concurrently with preceding computation. The exact indexer subsequently validates the prediction, and only missing KV caches require corrective reads. As a result, the corrective reads only consume 0.2--6.4\% of the time cost of append prefill. Model computation consequently account for 91.5--97.3\% of \system's append prefill time cost, compared with 62.1--81.5\% of \systembase. The portion of corrective reads varies across models and requests, leading to divergent speedup. For both DeepSeek-V4 variants, corrective reads occupy at most 1.4\% of the critical path, producing consistent speedups of 1.29-1.34$\times$. GLM-5.2 exposes 3.5-6.4\% time for corrective reads time and therefore shows a wider 1.13-1.49$\times$ speedup range. 

Prediction itself adds little overhead. It contributes only 0.3--1.2\% of the time cost of append prefill across the nine cases. As mentioned in \S\ref{sec:method_impl}, our implementation manages to overlap corrective reads and prediction with each other. However, when the prediction achieves high recall, the corrective reads may not be able to hide the prediction overhead. Thus, we place both of them on the critical path in Figure~\ref{fig:expr_case_studies}. Despite this, the prediction overhead is extremely small. Besides, we can also observe that the time cost of prediction is smaller than exact indexer, which verifies the effectiveness of our synchronization-free parallel prediction method.

\mysubsubsection{Serving efficiency}
\label{sec:expr_serving}
Although \system focuses on append prefill efficiency, we also conduct experiments under end-to-end serving. Specifically, we send the requests to serving frameworks following a Poisson process with a request arrival rate of 0.005, and report P95 TTFT and P95 TPOT in Figure~\ref{fig:expr_e2e_serving}.

In end-to-end serving, TTFT also involves the time cost of queuing, so the speedup range becomes narrower compared to the results in Figure~\ref{fig:expr_e2e_ttft}. Nevertheless, \system still outperforms Tutti by 1.05-1.51$\times$, LMCache by 1.31-3.04$\times$, and \systembase by 1.01-1.68$\times$ in terms of P95 TTFT. 

Meanwhile, the TPOT latencies of all competitors differ by approximately 3\%, which is reasonable since we do not alter the decode computation. Tutti slightly increases TPOT in some cases, which is because it packs the KV pages on GPUs before offloading them to SSDs and thereby leads to SM contention. In contrast, \system leverages CPU-assisted write packing to avoid competing SMs with model execution. In short, \system accelerates append prefill substantially while maintaining decode efficiency.

\begin{table}[!t]
\centering
\small
\caption{Ablation studies of our best-effort read packing and CPU-assisted write packing techniques (DeepSeek-V4-Flash, Trace 2). We show the relative degradation in P95 TTFT and TPOT compared with the setting in which both our read and write packing techniques are enabled.}
\label{tb:expr_ablation_packing}
\begin{tabular}{cccc}
\toprule
Read & Write & TTFT & TPOT \\
\midrule
No packing & CPU-assisted (ours) & 4.2\%$\downarrow$ & 0.3\%$\downarrow$ \\
Best-effort (ours) & No packing & 3.1\%$\downarrow$ & 5.5\%$\downarrow$ \\
Best-effort (ours) & GPU packing & 2.7\%$\downarrow$ & 3.7\%$\downarrow$ \\
\bottomrule
\end{tabular}
\end{table}

\subsection{Effectiveness of SSD I/O Optimization}
\label{sec:expr_ablation}

\mysubsubsection{Read and write packing}
We assess the effectiveness of our packing approaches that address the non-contiguous I/O issues. As shown in Table~\ref{tb:expr_ablation_packing}, when our best-effort read packing is disabled, the P95 TTFT latency is degraded by 4.2\%, showing that the performance of sparse SSD reads is still important even though we can reduce the I/O exposed on critical path. 
For SSD writes, we compare with two approaches. The first directly triggers non-contiguous GPU-SSD writes without packing (denoted as ``No packing''), whereas the second performs packing on GPUs and then triggers contiguous GPU-SSD writes (denoted as ``GPU packing''). It can be seen that both approaches degrade P95 TPOT performance by 5.5\% and 3.7\%, respectively. The GPU packing approach enables contiguous SSD writes but competes GPU SMs with model execution, while the no packing approach suffers from non-contiguous I/O. In contrast, our CPU-assisted packing approach facilitates contiguous I/O while avoiding SM contention. Besides, the degradation in TPOT also harms TTFT due to the read-write interference in SSDs.

\begin{figure}[!t]
\centering
\includegraphics[width=\linewidth]{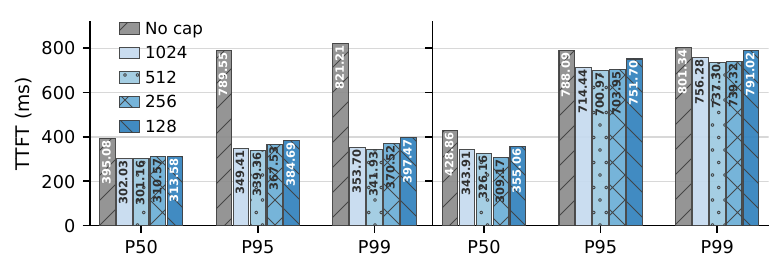}
\caption{Percentile TTFT latencies when setting different write capping limits (DeepSeek-V4-Flash, Traces 2 and 3).}
\label{fig:expr_sensitivity_write_cap}
\end{figure}

\mysubsubsection{Write bandwidth cap}
Figure~\ref{fig:expr_sensitivity_write_cap} evaluates the write bandwidth cap using percentile TTFT. Without write capping, it suffers from severe read-write interference, bumping P99 TTFT substantially compared with a 512 MB/s cap. Meanwhile, an overly tight cap also reduces serving efficiency. With a write cap of 128 MB/s, P99 TTFT is 16.2\% and 7.3\% higher than that of 512 MiB/s. This is because the limited write bandwidth slows down KV offloading, causing pending KV writes to accumulate and increasing queueing time. In contrast, \system consistently achieves sound TTFT performance with moderate caps. Thus we do not need to laboriously tune the write cap, demonstrating the robustness of our method.

\section{Related Work}
\label{sec:related}

\mysubsubsection{Hierarchical and SSD-backed KV storage}
Existing systems extend KV capacity through hierarchical GPU-CPU-SSD or even remote storage, often coupled with cache-aware scheduling and overlapped transfers~\cite{cachedattention,mooncake,memserve,lmcache,hicache,strata}. As discussed in \S\ref{sec:intro}, owing to the explosive growth of LLM demands, CPU DRAM has become more and more costly recently, which inspires us to focus on SSD-centric KV storage solutions. A line of research improves the GPU-SSD path through direct or KV-oriented storage abstractions~\cite{bam,geminifs,gofs,tutti}. These systems assume that a reusable prefix identifies all required KV caches before prefill, whereas sparse attention reveals a query-dependent subset only during layer execution. \system therefore coordinates sparse selection with SSD I/O instead of loading the full cached prefix.

\mysubsubsection{Sparse-attention LLMs and serving}
Research on sparse attention either retrofits dense models with token-, page-, or pattern-level selection~\cite{sparq,quest,minference}, or trains sparse selection into the model itself~\cite{nsa,moba,deepseekv32,deepseekv4,glm52}. Related mechanisms, including IndexCache and model-native IndexShare, reuse selection results across layers~\cite{indexcache,glm52}. \system does not alter these algorithms. Instead, it serves models with native sparse selection while preserving their exact outputs.

Several systems manage sparse KV placement between GPU and CPU memory without predicting future demand~\cite{sparseServe,hisparse}. Predictive systems instead speculate future KV selections to hide transfer latency~\cite{infinigen,freekv,dualdecoder,solidattention}. IMPRESS, InfiniGen, SparseServe, and SolidAttention derive sparse access from originally dense-attention LLMs and would change their outputs~\cite{impress,infinigen,sparseServe,solidattention}. Most predictive designs also target decode, where adjacent-step similarity enables prediction. ECHO~\cite{echo} supports prefetching for prefill of sparse attention, but it provides insufficient overlap for SSD latency, as discussed in \S\ref{sec:obs}. \system provides a longer lookahead for append prefill, corrects prediction misses before attention, and co-designs fragmented SSD reads with background KV writes.

\mysubsubsection{Agentic serving systems}
Agentic serving systems exploit request dependencies, tool lifetimes, or session locality to schedule programs and retain reusable KV caches~\cite{parrot,autellix,thunderagent,continuum,mori,cachewise,smetric,ampd,dualpath}. Workload studies show that agents repeatedly append to long contexts while producing relatively short outputs~\cite{tracelab,agenttokencost}. Orthogonal to these scheduling and placement policies, \system optimizes the inference path of each append prefill by discovering and fetching its query-dependent sparse KV caches from SSD.

\section{Conclusion}
\label{sec:conc}

In this work, we presented \system, a novel agentic serving framework for sparse attention LLMs with SSD-centric KV storage. We first characterized agentic workloads, revealing that the bottleneck of SSD reads occurs in append prefill and that read efficiency should be emphasized. Then, we developed an efficient lookahead prediction approach to speculatively prefetch KV caches while performing corrective reads to ensure lossless model outputs. We further tailored a series of SSD I/O optimization for sparse attention LLMs under agentic workloads. Empirical results show that \system achieves up to 1.57-3.69$\times$ (1.22-1.85$\times$ on average) in terms of the time to first token latency compared to existing works.

\bibliographystyle{plain}
\bibliography{references}

\end{document}